\documentclass[aps,twocolumn,a4paper,pra,10pt]{revtex4-1}
\usepackage{amsmath,graphicx}
\usepackage{braket}
\newcommand{\rf}{\ensuremath{\rm {\it rf}}}
\newcommand{\df}{\ensuremath{\delta\!f}}

\begin{document}

\title{Minimally-destructive detection of magnetically-trapped atoms using frequency-synthesised light}

\author{M. Kohnen}
\author{P. G. Petrov}
\author{R. A. Nyman}\thanks{To whom correspondence should be addressed: r.nyman@imperial.ac.uk}
\author{E. A. Hinds}

\affiliation{Centre for Cold Matter, Blackett Laboratory, Imperial College London, Prince Consort Road, London SW7 2AZ, United Kingdom}

\date{\today}

\begin{abstract}
We present a technique for atomic density measurements by the off-resonant phase-shift induced on a two-frequency, coherently-synthesised light beam. We have used this scheme to measure the column density of a magnetically trapped atom cloud and to monitor oscillations of the cloud in real time by making over a hundred non-destructive local density measurments. For measurements using pulses of $10^{4}-10^{5}$ photons lasting $\sim10\mu$s, the precision is limited by statistics of the photons and the photodiode avalanche. We explore the relationship between measurement precision and the unwanted loss of atoms from the trap and introduce a figure of merit that characterises it. This method can be used to probe the density of a BEC with minimal disturbance of its phase.

\end{abstract}

\maketitle

Ultracold atoms play a central role in modern metrology, matter-wave interferometry and many-body quantum physics. In all these applications one aims to detect the atoms with high efficiency and low noise. This can be achieved by resonant excitation, detecting either the absorption of the probe light or the corresponding induced fluorescence. Often, however, the cloud is optically thick. For example in a typical trapped Bose-Einstein condensate (BEC) the resonant optical absorption length is only 30\,nm or less. In addition each spontaneous emission destroys quantum coherence and promotes the loss of atoms from the trap. These problems can both be mitigated by detuning the light to make the sample optically thin and the scattering rate low.  Detection is then accomplished by measuring the phase shift that the atoms impose on the light. This dispersive detection has the advantage that it can be repeated many times on the same cloud. It is therefore sometimes called ``non-destructive''.

Non-destructive optical detection permits  multiple measurements of the same trapped atomic sample with repetition rates in excess of 100~kHz; faster than the natural timescales for density evolution in a trapped ultracold cloud set by the trapping frequencies~\cite{andrews96, petrov07}. Consequently the method can be used to explore dynamics in an atomic cloud or to monitor the time evolution of matter wave interference, e.g. in a BEC. With rapid readout, active feedback on the cloud becomes feasible, opening possibilities for cooling the motion~\cite{fischer02, kubanek09} or preparing novel quantum states.

The optical phase shift induced by the atoms is detected by interference with a reference beam preferably derived from the same laser in order to reject phase drift. The beams could be separated spatially, with one path going through the atom cloud, however, it is technically demanding to maintain adequate mechanical stability~\cite{petrov07}. Alternatively, separation in frequency allows them to travel on a common path while still accumulating a differential phase shift through the frequency-dependence of the atomic polarisability~\cite{bjorklund83}.

Previously, frequency modulation (fm) sidebands have been used to provide additional frequencies. Atoms dropped from a MOT have been seen using fm sideband detection with a current-modulated diode laser~\cite{savalli99}, and within a MOT, atoms have been probed using electro-optic modulation (EOM) to produce fm sidebands~\cite{lye99}. These fm methods produce (at least) three frequencies --- a carrier and two sidebands --- rather than two. One detection strategy is to tune all three frequency components above (or below) the atomic transition, then the sideband closest to resonance has the main phase shift and the strong carrier acts as the local oscillator. This has the drawback that it induces a light shift in the energy of the atoms. The shift can be eliminated by placing the sidebands symmetrically around the atomic resonance, but then the cloud is heated by the resonant carrier. In order to detect atoms in an optical lattice, Lodewyck \textit{et al.}~\cite{lodewyck09}, using an EOM, have suppressed the carrier by choosing a specific, high modulation index (2.4), but this has the effect of putting power into higher-order sidebands that contribute inefficiently to the signal.

In this paper, we show how an acousto-optical modulator (AOM) can synthesise the required two frequencies without any other sidebands. We couple this dual-frequency light into an optical fibre that allows easy delivery to a remote site where cold atoms are to be measured. We find that the relative phase of the two beams is exceedingly robust when transported in this way. Using a phase sensitive detector to read out both quadratures of the observed beat note, we study the phase noise and compare this with the expected noise floor due to Poisson statistics of the coherent laser light. We then apply this two-frequency interferometer to the dispersive detection of magnetically trapped $^{87}$Rb atoms. We measure the spectrum of the phase shift induced by the atoms and investigate the extent to which the measurement is non-destructive. To demonstrate the utility of the method, we non-destructively measure the centre-of-mass oscillations of a magnetically trapped atom cloud. Finally, we describe how to optimise the detection scheme, using a figure of merit based on sensitivity and destructiveness.

 All the methods described above are ultimately limited by the photon shot noise~\cite{horak03, lye03}, leading to the standard quantum limit. We note that there are also methods to go below that limit~\cite{caves81, kasapi98, mabuchi99, long07, bernon11}.

\section{The detector}
\subsection{Experimental setup}

\begin{figure}  
       \includegraphics[width=7.5cm]{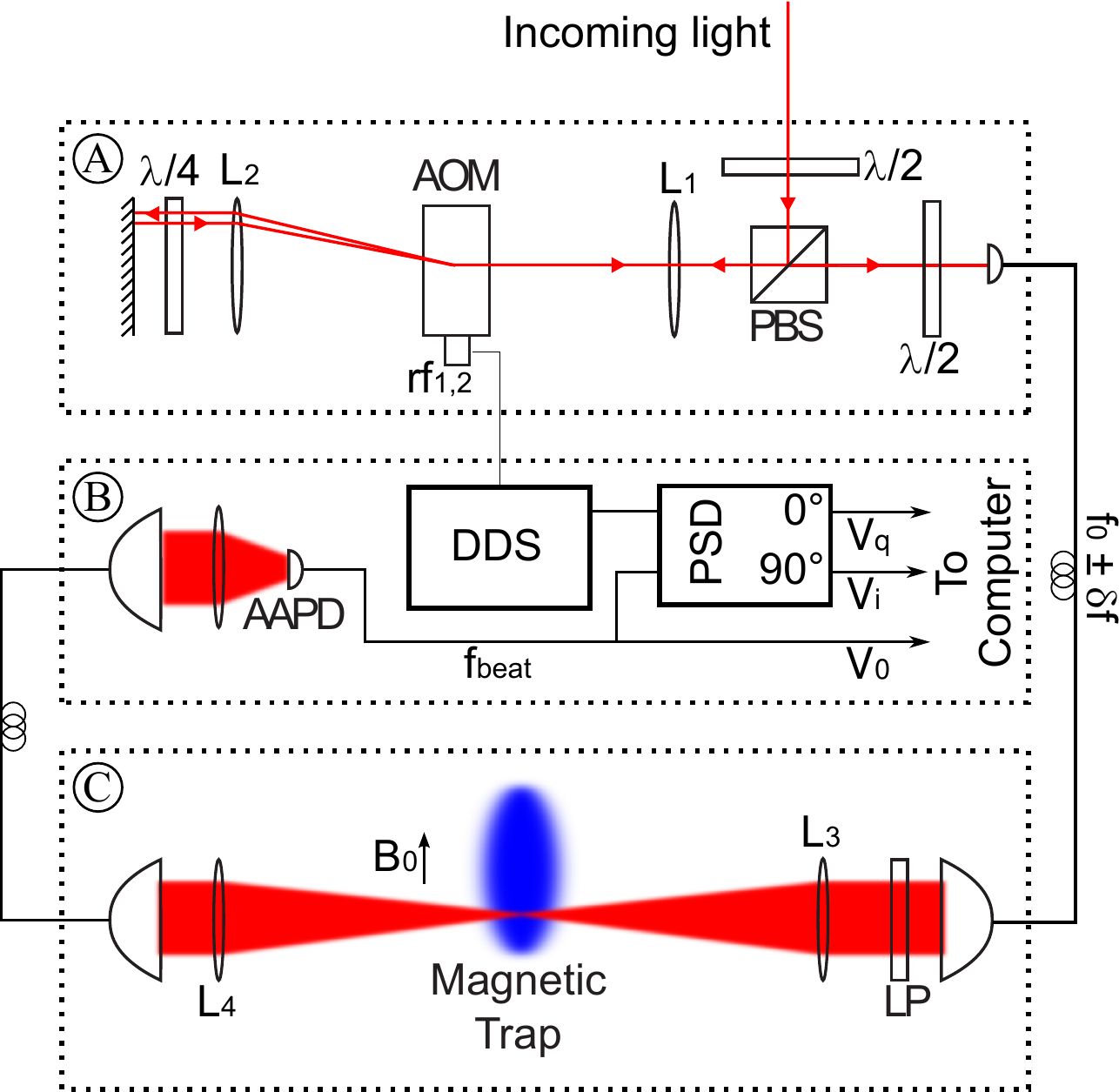}
        \caption{Experimental setup. (A) Synthesis: The incoming beam passes twice through an acousto-optic modulator (AOM) driven at two frequencies $\rf_1$ and $\rf_2$, The quarter-wave plate ($\lambda / 4$) rotates the linear polarisation so that the light is coupled out by a polarising beam cube (PBS). This light is coupled into a single-mode fibre. (C) Experimental chamber: The beam leaving the fibre is collimated, passed through a linear polariser (LP) and focussed ($L_3$) onto the magnetic trap before re-collimation ($L_4$) and coupling back into a multi-mode fibre. (B) Demodulation: Light is detected by an analogue avalanche photo-diode (AAPD) A phase sensitive detector (PSD) determines the in-phase and quadrature part of the beat note signal. Phase-stable modulation and reference signals are produced by the same multi-channel direct digital synthesizer (DDS).}
        \label{fig:setup}
\end{figure}

The setup for synthesising the two optical frequencies is shown in figure \ref{fig:setup}(A). A collimated, linearly polarised laser beam enters the input and passes through a half wave plate ($\lambda/2$) that adjusts the polarisation to be vertical. This light is directed by the polarising beam splitter (PBS) through lens $L_1$, acousto-optic modulator (AOM), and lens $L_2$ which re-collimates it. Retro-reflection through the quarter-wave plate rotates the linear polarisation to horizontal. After passing again through $L_2$, AOM and $L_1$, the light is transmitted by the PBS. A final $\lambda/2$ rotates the polarisation to any desired angle before the light is coupled into a single-mode fibre for transmission to the atom cloud.

The AOM is driven at two rf frequencies, $\rf_1$ and $\rf_2$, produced by a 4-channel direct digital synthesizer DDS (Novatech DDS 409B) and passively summed.  The central beam emerging from the PBS contains the two desired frequencies, let us call them $f_0 \pm \df$, produced by the double-pass AOM shifts $2\,\rf_{1}$ and $2\,\rf_{2}$. On either side of this are beams at unwanted frequency $f_0$, produced by a shift of $\rf_1$ in one pass and $\rf_2$ in the other. The use of long focal length ($400 ~ \rm{mm}$) lenses ensures that these side beams are well resolved from the main beam, so that less than $1\%$ of their power is coupled into the fibre when $\left|2\,\df \right| > 20 ~ \rm{MHz}$. The two rf amplitudes are adjusted to balance the power of the two desired frequency components in the light.

The single-mode fibre guiding the light to the experimental chamber, figure \ref{fig:setup}(C), enforces the best possible spatial mode overlap of the two frequency components. The light leaving this fibre is collimated, linearly polarised (LP) and then focused to a waist of $w_0 = 55 ~ \rm{\mu m}$ by lens $L_3$ of focal length $250 ~ \rm{mm}$. After passing through part of the magnetically trapped cloud of atoms, the beam is re-collimated and coupled into a multi-mode fibre, which transports up to $90 \%$ of the light collected by $L_4$ to the detection electronics shown in figure \ref{fig:setup}(B).

The light is detected with $77\%$ quantum efficiency by an analogue avalanche photodetector (AAPD)\footnote{Analog Modules Inc. model 712A-4, consisting of a PerkinElmer C30902E avalanche photodiode followed by a transimpedance amplifier. We are informed that the model is obsolete, but that replacement part will be available soon, and has equivalent or better specifications.} which produces a voltage proportional to the rate of detected photons. With equal power at frequencies $f_0\pm\df$, this rate may be written as $R+R \cos(\Omega t+\phi)$, where $\Omega=4\pi\df$ is the beat angular frequency and $\phi$ is the relative phase between the two frequency components. We set $\Omega$ to $2\pi\times60\,\rm{MHz}$ in the experiments presented here. We detect the in-phase ($V_i$) and quadrature ($V_q$) components of the beat using a phase sensitive detector (PSD) whose local oscillator at angular frequency $\Omega$ is generated by the same DDS that drives the AOM. The outputs of the PSD are low-pass filtered at $650\,$kHz, then digitised using a 14-bit analogue-to-digital converter card NI \hbox{PCI-6133} with 1.3~MHz analogue bandwidth.

\subsection{Noise}
\label{sec:noise}

\begin{figure}  
	\includegraphics[width=8.5cm]{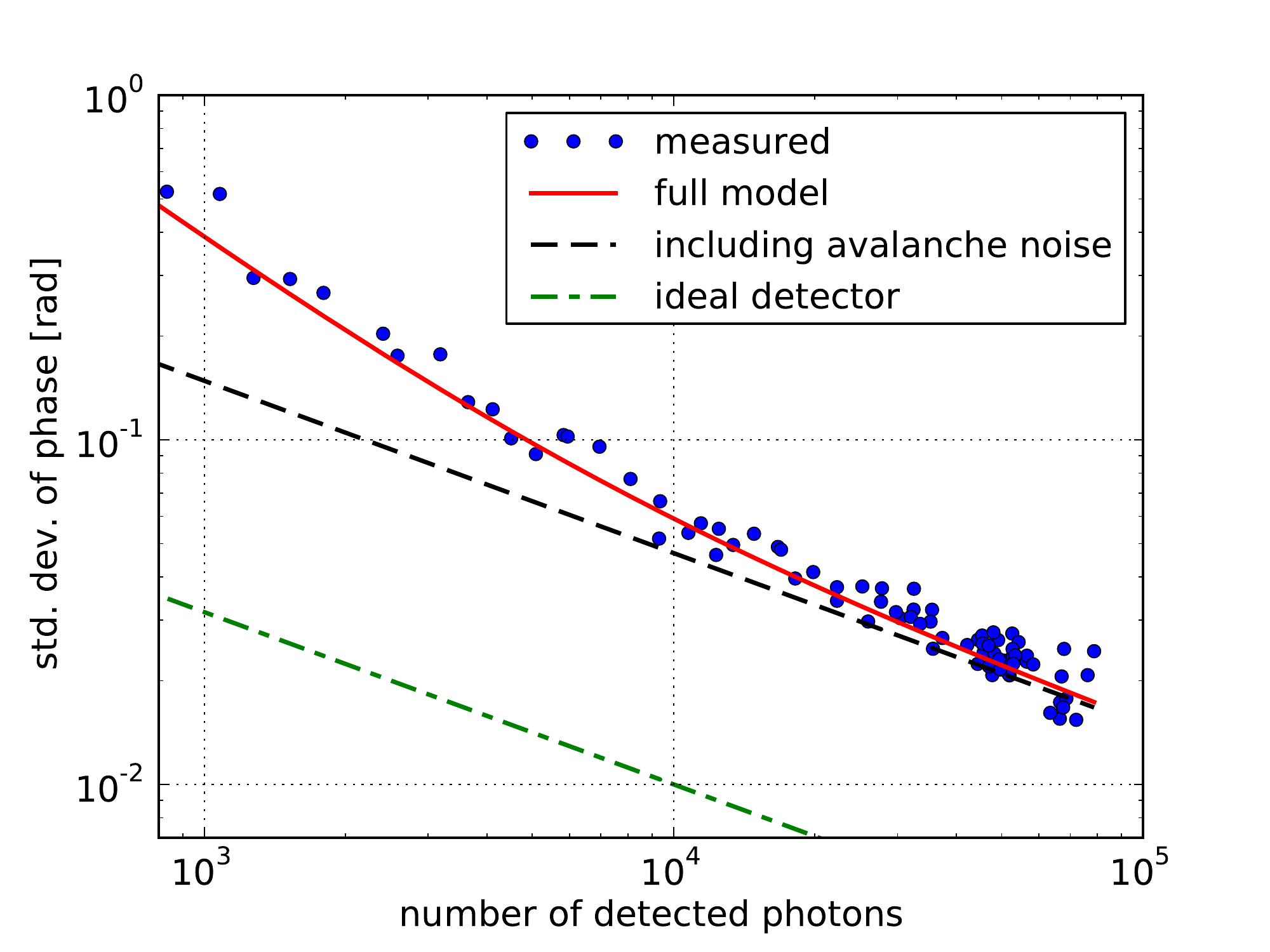}
	\caption{Standard deviation of the beat-note phase versus the number of photons detected in an integration time of $10 ~ \rm{\mu s}$. Dash-dot line: ideal detector having no avalanche noise. Dashed line: analogue detector with avalanche. Solid line: full noise including that of the electronics. Dots: measured noise levels.}
        \label{fig:noise}
\end{figure}

The digitised PSD outputs are integrated over a measurement time $T$ which contains an integer number of beat cycles. During this time, the mean number of detected photons is $N=R T$ and the standard deviation of this number due to shot noise is $\sqrt{N}$. When $\phi=0$ the beat note is in phase with the PSD local oscillator and the signal $V_i$ corresponds (through the various amplifier gains) to a count rate $R \cos^2(\Omega t)$ which integrates to a time average of $\frac{1}{2}N$. More generally, the time average $\overline{V_i}$ (or $\overline{V_q}$) corresponds to a count of $\frac{1}{2}N\cos \phi$ (or $\frac{1}{2}N\sin \phi$). The noise power is equally distributed between the two outputs of the PSD, corresponding to standard deviations in the count of $\sqrt{N/2}$ each. In our experiments, the standard deviation of the phase due to this shot noise is small because $N\gg 1$, leading to a value $\sigma_\phi=\frac{\sqrt{N/2}}{N/2}=\sqrt{2/N}$. This level of noise is shown by the dash-dotted line in Fig.~\ref{fig:noise} versus the number of detected photons.

Our instrument is not expected to reach the shot noise level because the photodiode current acquires additional noise as a result of the avalanche process~\cite{teich86}, giving a larger phase uncertainty
\begin{equation}\label{eq:sigmaPhi}
\sigma_\phi=X\sqrt{2/N}
\end{equation}
We have measured this extra noise factor $X$ and find that it has the value $3.3\pm 0.3$ (somewhat higher than the 2.2 indicated in the manufacturer's specifications). The correspondingly higher noise level is shown by the dashed line in Fig.~\ref{fig:noise}. In addition, the photodiode amplifier and the PSD contribute electronic noise, which is independent of the number of photons detected and therefore appears as a further phase uncertainty proportional to $1/N$. When this is added in quadrature to the other noise we obtain the total anticipated noise, indicated by the solid line in Fig.~\ref{fig:noise}.

The points in Fig.~\ref{fig:noise} show the phase noise that we have measured. In a single phase measurement we detect a light pulse lasting $T=10~\mu$s and we integrate the PSD outputs to obtain mean values $\overline{V_i}$ and $\overline{V_q}$, from which we determine a phase $\arctan(\overline{V_i}/\overline{V_q})$. From the standard deviation of 50 such phase measurements we determine one point on the graph in Fig. 2. The number of detected photons is determined directly from the mean AAPD signal, $V_0$. This procedure is repeated over a wide range of light intensities to produce the set of data points. These measurements show that the noise in our instrument is well understood and that there are no other significant noise sources.

The electronic noise becomes increasingly important as the count rate is reduced, and is equal to the avalanche-degraded shot noise at 580 detected photons/$\rm{\mu s}$. There is scope for suppressing the electronic noise of the PSD, in which case the photon rate could be reduced to $\sim 200$ photons/$\rm{\mu s}$ before the noise reaches that of the AAPD electronics. If the analogue detection were replaced by pulse counting, our instrument could enjoy the noise indicated by the dash-dotted line in Fig.~\ref{fig:noise}. However, currently available pulse-counting APDs are limited by detection dead time to less about 10~photons/$\rm{\mu s}$, which would slow down the precise measurement of phase.

There can also be systematic noise in phase of the beat note due to fluctuations in the difference of optical path lengths for the two frequency components. Most important in this regard is the region between the AOM and the retro-reflection mirror (see part A of Fig.~\ref{fig:setup}), where the two beams take different paths. After enclosing this part of the apparatus to shield it from air currents, the optical path fluctuations were too small to see as excess noise in Fig.~\ref{fig:noise}. On increasing the number of detected photons to a million, by increasing the integration time to $T=100~\mu$s, we found an excess $\sigma_{\phi}$ of 2\,mrad. Once in the optical fibre, no more optical path length noise is observed; neither shaking the fibre nor changing its length between 2 and 20~m induces noise. The mean phase drifts by a radian over tens of minutes, presumably because of mirror movement in the same sensitive region of the apparatus. In principle, this technical noise can be removed by using active feedback to stabilise the phase, but we have not done so here because the noise is small and the drifts are slow.

\section{Atom-light interaction}
\label{sec:theory}

The light used in our experiment is tuned near the $\ket{F=2,m}\rightarrow \ket{F'=3,m'}$ hyperfine transition of the $D_2$ line of $^{87}\rm{Rb}$ and propagates perpendicular to the magnetic field axis (see figure \ref{fig:setup}, part C). Almost all the atoms in the magnetic trap are prepared in the $\ket{F,m} = \ket{2,+2}$ state, and we will assume for the moment that they remain there (negligible optical pumping). If the light travels through an atomic vapour of column number density  $\rho_a$ the phase shift $\theta$ and fractional attenuation $\alpha$ of the optical field are, in the limit of negligible saturation,
\begin{equation}
\theta(f) =\frac{7}{2} \left( \sum_{m'=1}^3 p_{m'} S_{m'} \frac{\gamma \delta_{m',f}}{\delta_{m',f}^2 + \gamma^2} \right) \frac{3\lambda^2 \rho_a}{2 \pi}
\label{eq:theta}
\end{equation}
\begin{equation}
\alpha(f) =\frac{7}{2}\left( \sum_{m'=1}^3 p_{m'} S_{m'} \frac{ \gamma^2}{\delta_{m',f}^2 + \gamma^2} \right) \frac{3 \lambda^2 \rho_a}{2 \pi},
\label{eq:alpha}
\end{equation}
where $p_{m'}$ denotes the fraction of light power at frequency $f$ polarised to drive the atomic transition $\ket{m=2}\rightarrow\ket{m'}$, and
\begin{equation*}
S_{m'} = \left( \begin{matrix} 3 && 1 && 2 \\ -m' && m'-2 && 2  \end{matrix} \right)^2
\end{equation*}
is the square of the Wigner-3$j$ symbol. The detuning of the light frequency $f$ from resonance, $\delta_{m',f}$, includes the Zeeman shift of the transition frequency due to the magnetic field. The damping rate $\gamma$ is half the spontaneous decay rate of the upper level, and $\lambda$ is the wavelength.

Our apparatus measures the relative phase shift \hbox{$\phi = \theta(f_0+\df) - \theta(f_0-\df)$} between the two frequency components and the fractional change in their detected beat amplitude \hbox{$\epsilon \approx \left[\alpha(f_0+\df) + \alpha(f_0-\df)\right]$}. The latter equation holds because $\alpha\ll 1$. The fraction of light power scattered is $2\epsilon$.

\begin{figure}  [t]
       \includegraphics[width=8.5cm]{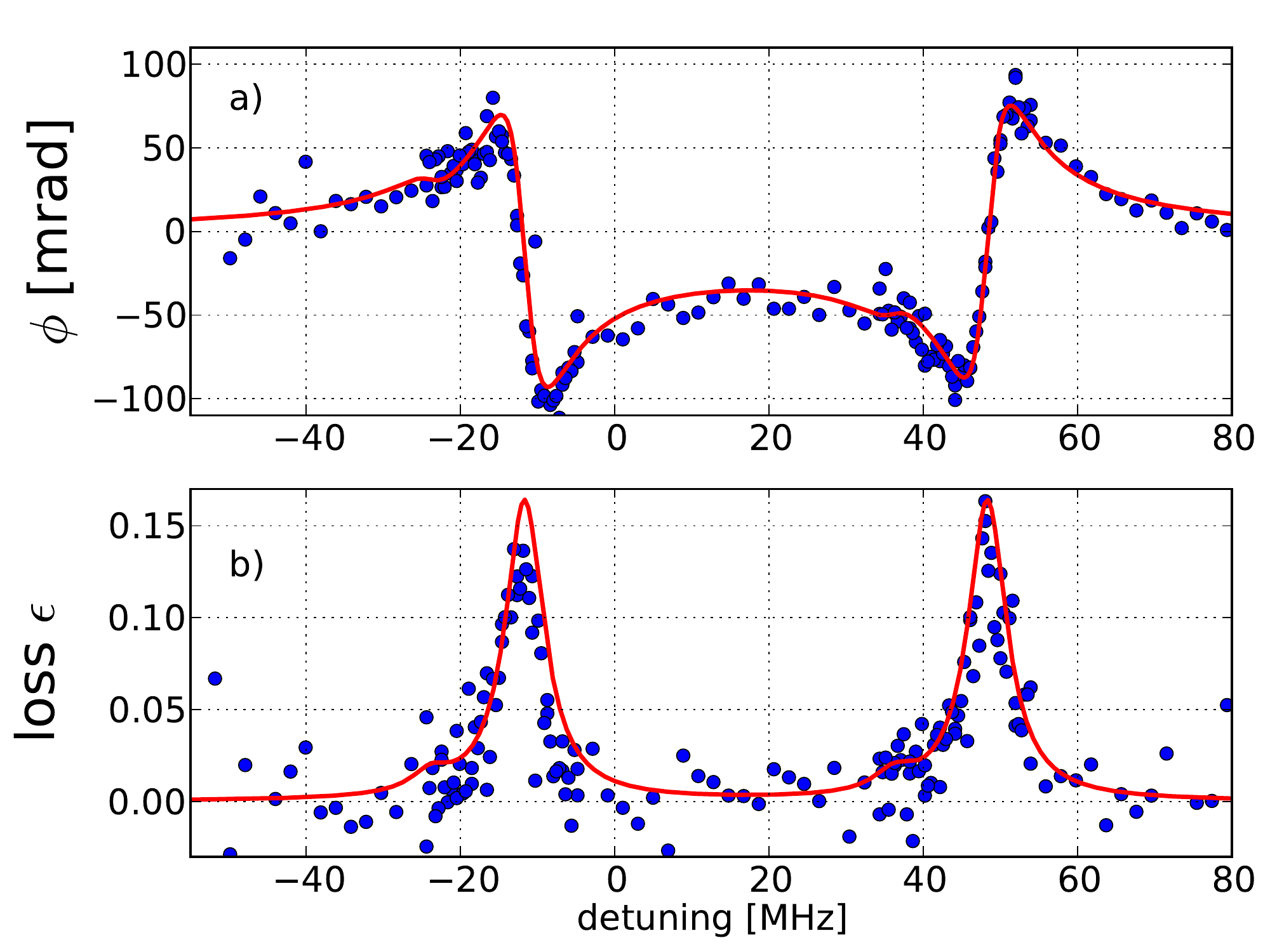}
        \caption{Phase shift and attenuation of $60 ~ \rm{MHz}$ optical beat note due to interaction with atoms in a magnetic trap. The light is polarised perpendicular to the magnetic field.  The abscissa is the detuning of the central laser frequency $f_0$ from an arbitrary zero. Dots: data measured in 32 shots, half with atoms and half without, each integrated over $10\,\mu$s. (a) Phase shifts $\phi$. Solid lines: least squares fit to Eq.(\ref{eq:theta}). The fit gives a column number density of $\rho_a = \left( 2.2 \pm 0.6 \right) \times 10^{12} ~ \rm{atoms~m^{-2}}$. (b) Fractional change $\epsilon$ in beat amplitude. The line uses the fit parameters from (a).}
        \label{fig:spectrum}
\end{figure}

\subsection{Detection of magnetically trapped atoms by phase-shift and absorption}
\label{sec:measurements}

The experimental chamber is supplied with a stream of cold $^{87} \rm{Rb}$ atoms from a low velocity intense source (LVIS) \cite{lu96}. Over a few seconds, these are captured in the main vacuum chamber by a U-MOT \cite{wildermuth04},  then transferred into a cigar-shaped magnetic trap produced by current-carrying wires. The magnetic field has a minimum of $B_0=0.6 ~ \rm{mT}$ at a distance of $2.5 ~ \rm{mm}$ from the surface of the central wire, and the trap has frequencies of $75 ~ \rm{Hz} ~ \times ~ 21 ~ \rm{Hz}$. Typically we load $2.4$ million atoms at a temperature of $60~\mu$K.

Figure \ref{fig:spectrum}(a) shows the phase shift $\phi$ for light polarised perpendicular to the magnetic field $ B_0$. The abscissa is the central frequency $f_0$ of the light relative to an arbitrary zero. Each phase shift is determined from the time-average $\overline{V_i}$ and $\overline{V_q}$ over a $10\,\mu$s detection window, corresponding to $\sim3 \times 10^5$ detected photons. The average phase $\phi=\braket{\arctan(\overline{V_q}/\overline{V_i})}$ is determined from 16 such measurements, spaced by 1\,ms and taken with atoms in the trap. The atoms are then released and the measurement is repeated to determine the background phase, which we subtract. The same data are used to determine the fractional change in beat amplitude $\epsilon$ by comparing the averages $\left\langle \sqrt{\overline{V_i}^2+\overline{V_q}^2} \right\rangle$ with and without atoms. These values are plotted in Fig.~\ref{fig:spectrum}(b).

The solid curve in Fig.~\ref{fig:spectrum}(a) shows a least-squares fit of Eq.~\ref{eq:theta} to the phase data, with column density and a central frequency offset as variable parameters. We see a dispersion feature when either of the two frequency components tunes through the $\ket{m=2}\rightarrow\ket{m'=3}$ resonance. This is due to the $\sigma^+$ component of the light. The $\sigma^-$ transition $\ket{m=2}\rightarrow\ket{m'=1}$, being fifteen times weaker, is not seen in the data. The fit gives a column density of $\rho_a=2.2 \times 10^{12}~\rm{atoms~m^{-2}}$, which is consistent with the density measured by destructive absorption imaging on a camera. Using the same fit parameters in Eq.~\ref{eq:alpha}, we obtain the line plotted in Fig.~\ref{fig:spectrum}(b), in good agreement with the observed variation in the amplitude of the beat note.

\begin{figure}  [t]
       \includegraphics[width=8.5cm]{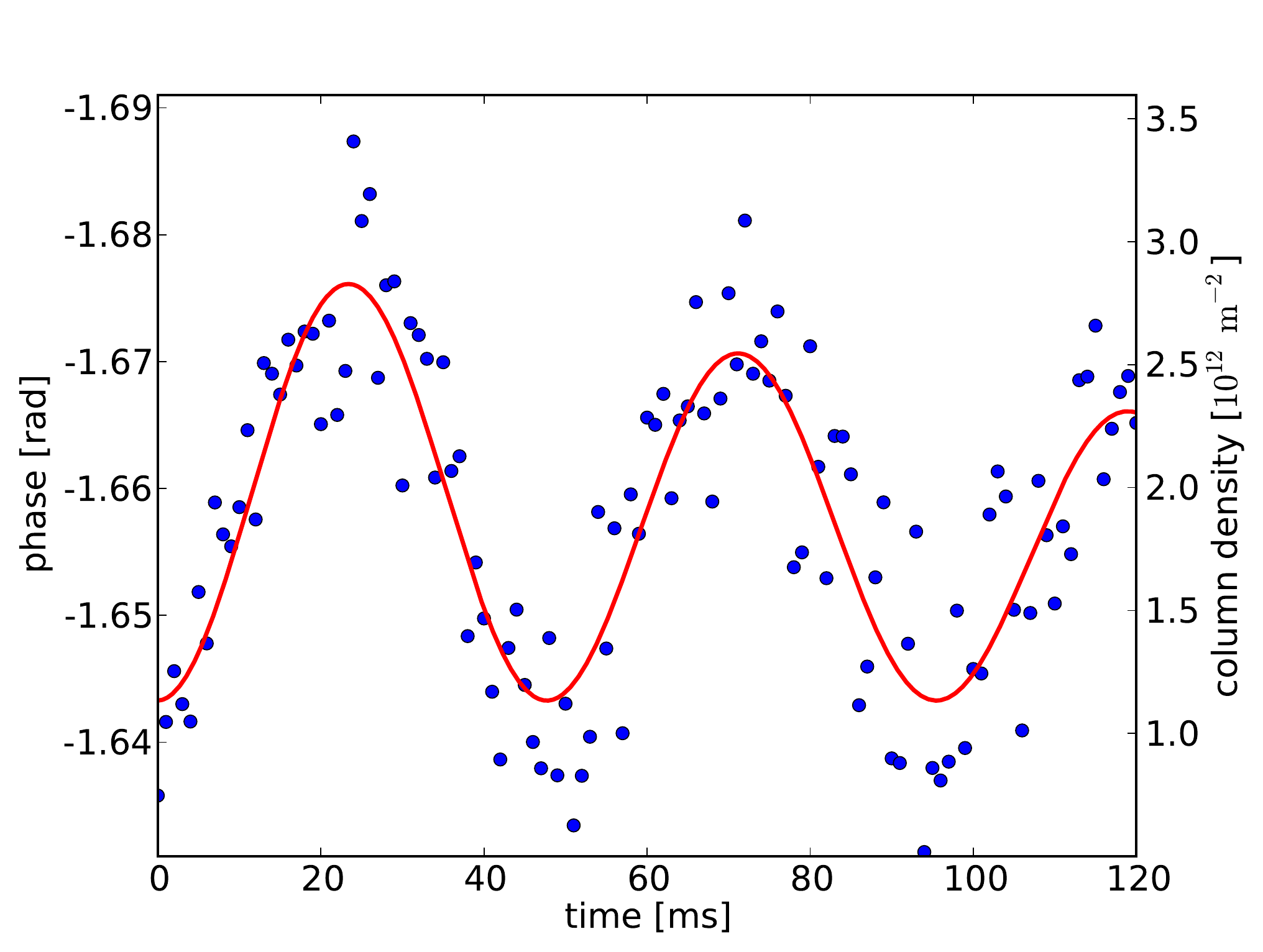}
        \caption{Centre-of-mass oscillations of a trapped atom cloud, observed through the modulated phase shift of the beat note in a single realisation of the experiment. The laser detuning corresponds to the frequency $17.3\,\rm{MHz}$ in Fig.~\ref{fig:spectrum}. Dots: $120$ phase shift measurements spaced $1\,\rm{ms}$ apart, each using a light pulse comprising $4 \times 10^5$ photons and lasting $50 ~ \rm{\mu s}$. Line: a fit, with trap frequency measured to be $21\pm 1$~Hz. After the measurement, less than 3\% of the atoms have been lost due to the probe light. The damping of the oscillations is due to anharmonicity of the trap.}
        \label{fig:single_shot}
\end{figure}

In Fig.~\ref{fig:single_shot} we show how the optical phase shift can be used to monitor the density evolution of a cold atom cloud. Here we have made measurements at 1\,ms intervals over a period of 120\,ms to record the centre-of-mass oscillation of atoms held in the magnetic trap. The optical phase shift is made sensitive to the motion by placing the probe beam approximately one cloud radius ($\sim 600\,\mu$m in this case) from the centre of the trap. For the purpose of this demonstration we set the cloud oscillating by making an intentional misalignment when we load it. This method allows us to determine the frequency and amplitude of the motion in a very short time, leaving the cloud largely undisturbed at the end of the measurement. By comparison, the same measurement using the standard method of resonant absorption imaging would require several hundred seconds to perform because the cloud is destroyed by a single laser shot and has to be replaced by a fresh cloud each time. Since it is a routine part of any experiment on magnetically trapped atoms to measure the trap frequency and to null the trap oscillations from time to time, our methods is of practical value. Moreover, it opens the possibility of tracking dynamical evolution of the cloud in real time and of doing so with high spatial resolution and bandwidth. Rapid, non-destructive, local monitoring would make it possible to implement recent ideas of fast feedback and control~\cite{szigeti09,szigeti10}. This could be extended to many channels using the array of atom-photon junctions~\cite{kohnen11} that we have developed.

As shown by Eqs.~(\ref{eq:theta}) and (\ref{eq:alpha}) the atoms cannot induce an optical phase shift without also scattering the light. Each scatter imparts a recoil momentum to the atom, which heats the cloud, and also opens the possibility that the atom may be optically pumped out of the $\ket{m=2}$ state. Both of these effects reduce the density of the cloud. However, the scattering during this measurement is so weak that it cannot be responsible for damping the oscillation in Fig.~\ref{fig:single_shot}. We have confirmed this by allowing the cloud to oscillate in the dark and measuring at later times. We believe that the damping is in fact due to the anharmonicity of the trap. It is nonetheless relevant to consider at what level atoms \textit{are} lost from the trap as a result of the measurement, and this is the subject of the next section.

\subsection{Losses}
\label{sec:loss}
\begin{figure}  
       \includegraphics[width=8.5cm]{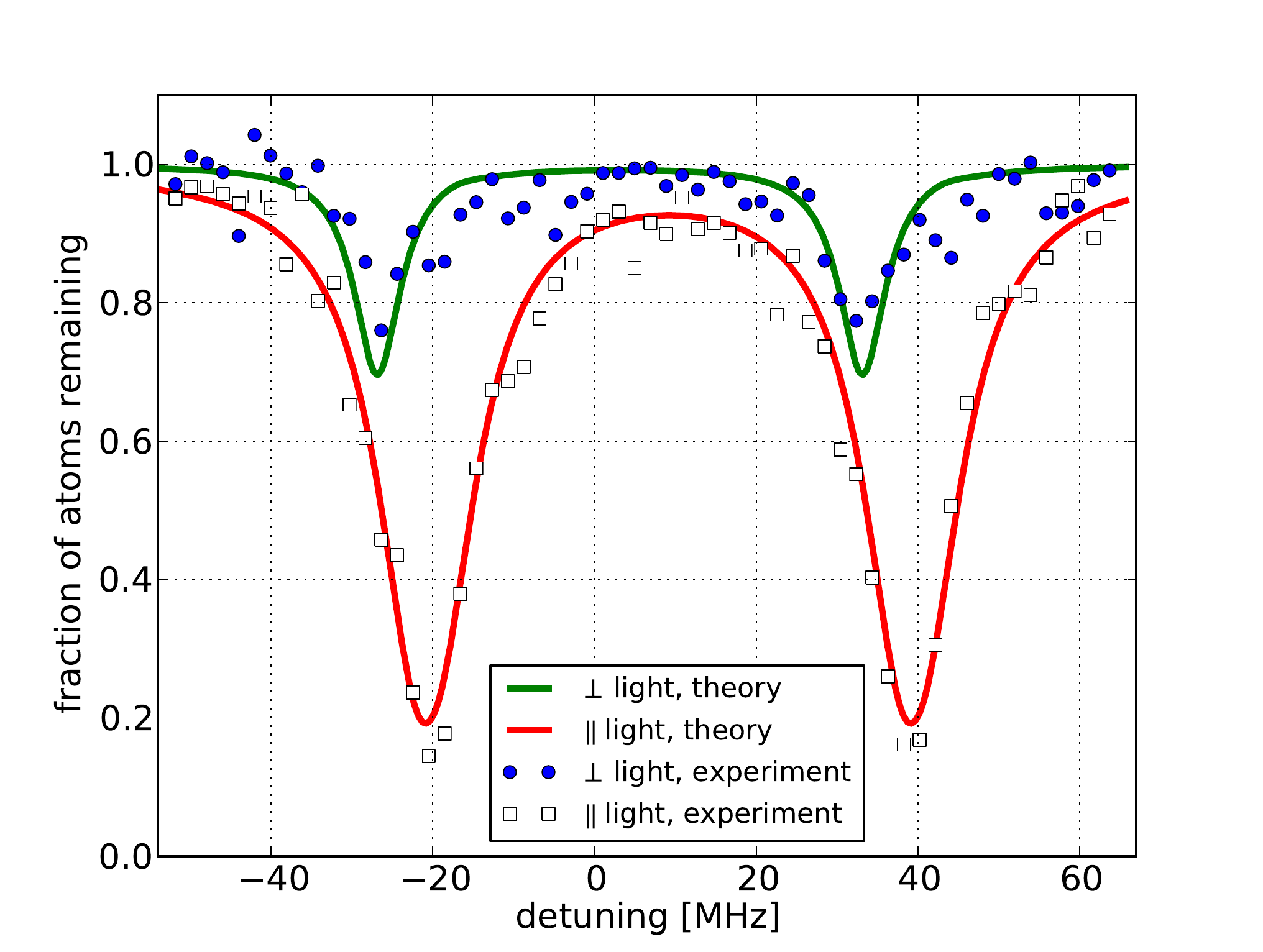}
        \caption{Fraction of atoms remaining in a magnetic trap after a measurement with $200$ pulses. Loss is mostly due to atoms being pumped into a weakly-trapped state, where their thermal energy is sufficient to take them out of the trap. This pumping effect is much stronger for light polarised parallel to the magnetic field than perpendicular. The solid lines are theoretical predictions from a rate equation model following the internal states of the atoms during the light pulses.}
        \label{fig:loss}
\end{figure}
\begin{figure}  
       \includegraphics[width=5.5cm]{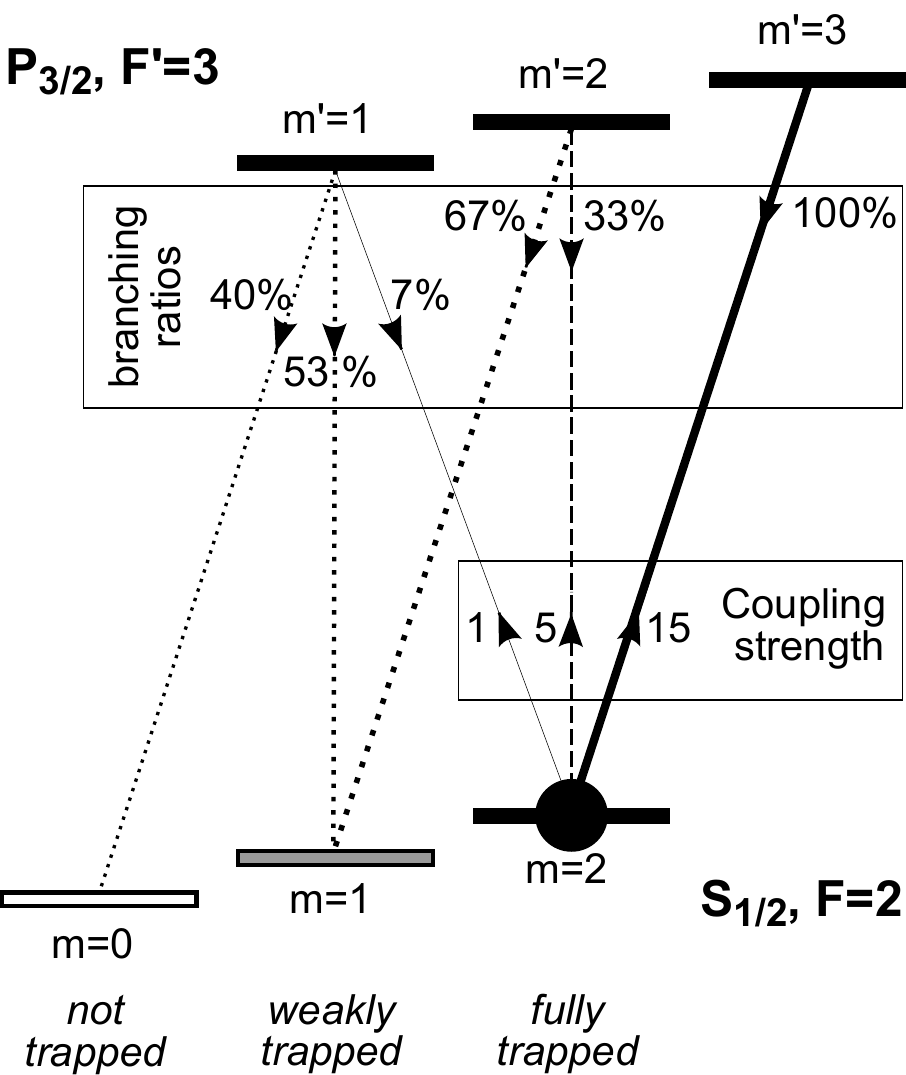}
        \caption{Relative coupling strengths for exciting trapped atoms in the $|F,m \rangle = |2,2\rangle$ state. Transitions driven by light perpendicular (parallel) to the magnetic field are indicated by a solid (dashed) line. Branching ratios for decay back to the ground states are also shown. Atoms pumped to state $\ket{2,0}$ are untrapped and therefore lost. Most of the atoms pumped to the weakly-trapped state $\ket{2,1}$ are also lost. Those remaining in state $\ket{2,2}$ are trapped.}
        \label{fig:transitions}
\end{figure}

We have measured the loss of atoms directly. The trapped atoms were illuminated by $200$ pulses of probe light, each lasting for $30 ~ \rm{\mu s}$, then the trap was switched off. After a further delay of  3~ms the number of atoms remaining in the cloud was measured by resonant absorption imaging on a CCD camera. The data points in Fig.~\ref{fig:loss} show the fraction of atoms remaining in the trap as a function of laser frequency. There are two curves. The one with weaker loss is measured with light polarised perpendicular to the magnetic field at the centre of the trap and with $6\times 10^5$ photons per pulse passing through the cloud. We see resonant loss dips caused by the $\sigma^-$ excitation to $\ket{m'=1}$. As shown in Fig.~\ref{fig:transitions}, this causes atom loss through its decays to the weakly trapped $\ket{m=1}$ state and to the untrapped $\ket{m=0}$, however, this excitation is 15 times weaker than the $\sigma^+$ cycling transition.  The data series showing stronger loss is measured using light polarised parallel to the trap field with $9\times 10^5$ photons per pulse. Here the loss is through faster excitation to $\ket{m'=2}$ state, followed by its preferred decay to the weakly trapped $\ket{m=1}$ state.

The curves in Fig.~\ref{fig:loss} are calculated using the following simple model. Atoms start in the strongly-trapped $\ket{m=2}$ state and we use rate equations to calculate how the populations of the levels in Fig.~\ref{fig:transitions} evolve over a single probe pulse. At the end of the pulse, atoms having $m=0, -1, \mbox{or} -2$ are discarded as untrapped, while those with $m=2$ are trapped. Of those in state $\ket{m=+1}$, we estimate that only 10\%  remain trapped because the gravitational force greatly lowers the barrier for escape. All trapped atoms are detected with equal efficiency in the absorption image because those having $m=1$ are quickly pumped into the $\ket{m=2}$ state.

After each probe pulse there is a delay of almost 1\,ms before the next pulse arrives. During this time the optically-pumped atoms move out of the probe region and are replaced by a new set of atoms. To a good approximation these are all in state $\ket{m=2}$. If we write the fraction of trapped atoms in the probe beam as $p$ and the fraction of these that are  pumped to untrapped states by one pulse as $q$, then the survival probability after one pulse is $1-q p$. Hence the fraction of trapped atoms surviving $k$ pulses is $\left( 1 - q p\right)^k $.

The solid lines in Figure \ref{fig:loss} plot the results of this model applied to our experiment, where the probe beam size in the cloud is $w=100\,\mu$m and $p=1.2\%$. The good agreement with experiment shows that, despite its simplicity, this model reliably relates the loss of atoms to the scattering rates.

With perpendicular polarisation, the optical pumping probability after one pulse is small enough that the following even simpler model suffices. Taking  $N_{\gamma}$ incident photons, the total number of $\sigma^-$ excitations per atom is $\epsilon_{1} N_{\gamma}$, where
\begin{equation}\label{eq:epsilon1}
\epsilon_1=\frac{\lambda^2}{40 \pi A}\left(\frac{\gamma^2}{\gamma^2+\delta_{1,f_0+\delta f}^2}+\frac{\gamma^2}{\gamma^2+\delta_{1,f_0-\delta f}^2}\right).
\end{equation}
This follows directly from Eq.~\ref{eq:alpha} with $p_1=\frac{1}{2}$ and $S_1=\frac{1}{105}$. Also, $\rho_a$ is replaced by $1/A$, where the beam area $A$ is related to the waist size $w$ by $A=\frac{1}{2}\pi w^2$. The excited state $\ket{m'=1}$ decays with 40\% probability to the untrapped state $\ket{m=0}$ and with 53\% probability to $\ket{m=1}$, which is lost 90\% of the time. Thus the probability that an atom in the probe volume will be lost as a result of the probe pulse is
\begin{equation}\label{eq:pq}
 q=0.88\epsilon_{1} N_{\gamma}.
\end{equation}
For perpendicular polarisation, the loss spectrum given by this approximation is almost indistinguishable from the theoretical curve in Fig.~\ref{fig:loss}. For parallel polarisation, however, the optical pumping is too strong for this approximation to suffice.

\subsection{Figure of merit}
\label{sec:fom}

\begin{figure} [t] 
       \includegraphics[width=8.5cm]{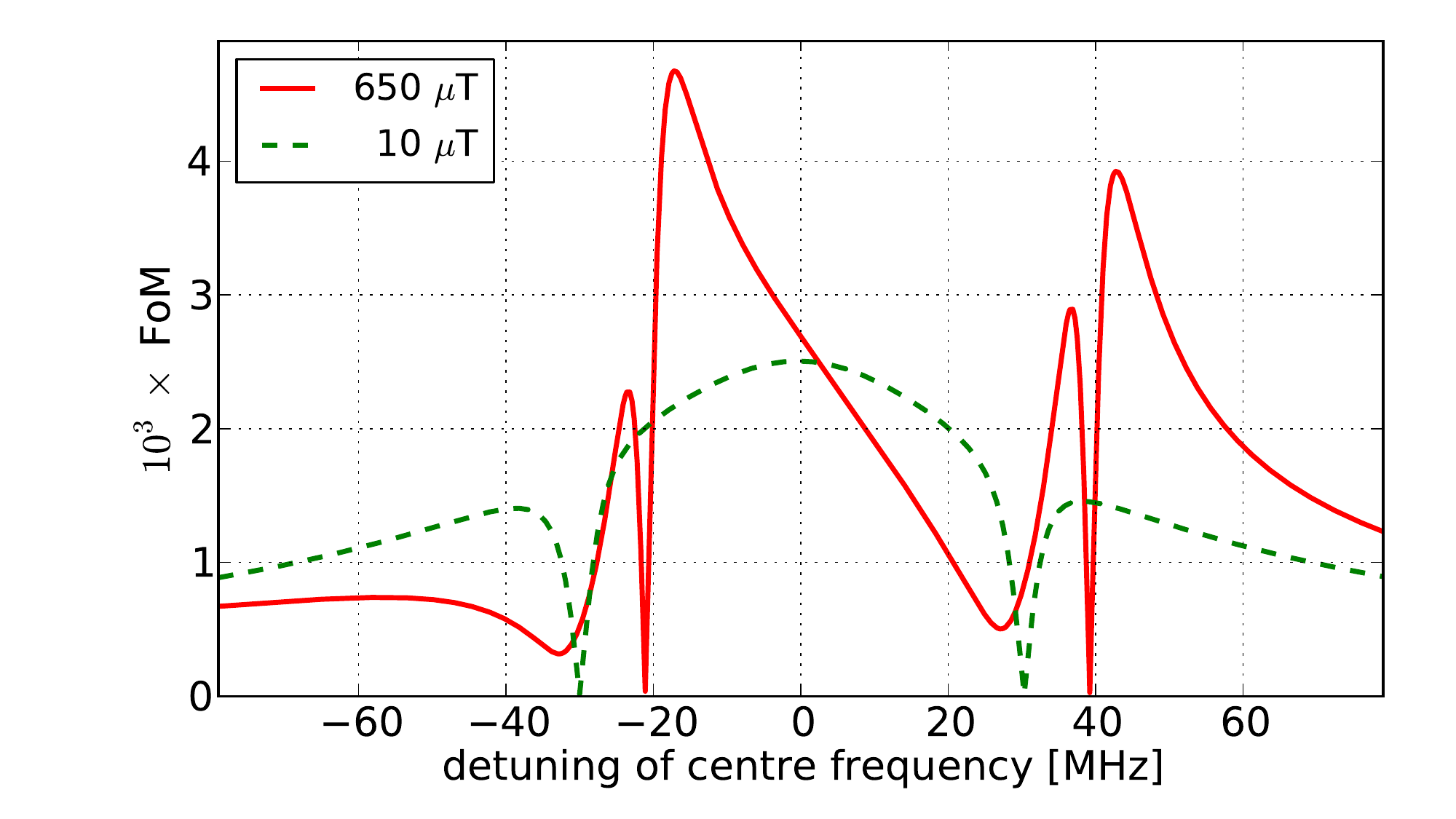}
        \caption{ Figure of merit (defined in text) for detecting $^{87}$Rb atoms using light with perpendicular polarisation. The abscissa shows detuning of the mean frequency $f_0$ from the $F=2\rightarrow F'=3$ cycling transition. The two light frequencies are separated by $60\,$MHz. Dashed line: $10\,\mu$T magnetic field. Solid line: $650\,\mu$T magnetic field.}
        \label{fig:FoM}
\end{figure}

The detection scheme presented in this paper is developed for multiple measurements of the atomic density on the same sample, as for example shown in figure \ref{fig:single_shot}. The aim is to infer the number of atoms in the probe beam $N_a$ from the measured phase $\phi$ with the smallest uncertainty $\sigma_{N_a}$. Since $\phi\propto N_{a}$ , the uncertainty is $\sigma_{N_a}=\frac{\sigma_{\phi}}{|\phi_{1}|}$. Here $\phi_{1}=\phi/N_a$ is the phase shift per atom in the light beam, given by Eq.~(\ref{eq:theta}) on replacing $\rho_a$ by $1/A$. It is also desirable to minimise $q$, the fraction of atoms lost from the probe region as a result of the probe pulse. We therefore define the figure of merit for the case of perpendicular polarisation as
\begin{equation}\label{eq:FoM}
FoM = \frac{1}{\sigma_{N_a}\sqrt{q}}\simeq \frac{|\phi_{1}|}{X\sqrt{2\epsilon_{1}}}\sqrt{\frac{N}{N_{\gamma}}},
\end{equation}
where the last step above makes use of Eqs.~(\ref{eq:sigmaPhi}) and (\ref{eq:pq}) and we take $\sqrt{0.88}\simeq1$. The factor $N/N_{\gamma}$ is the fraction of photons in the pulse that are detected. For our detector this quantum efficiency is 0.77. This figure of merit has the virtue that it does not depend on the number of atoms or the number of photons. It can be seen as the signal-to-noise ratio per atom, when the atoms have a $1/e$ probability of surviving the measurement. If the area of the beam is reduced, the FoM increases as $A^{-1/2}$, which motivates efforts to probe atom clouds using beams of small cross section \cite{horak03} \cite{kohnen11}.

The dashed line in Fig.~\ref{fig:FoM} shows how the figure of merit varies with the detuning of the central frequency $f_0$ from the atomic resonance frequency. We take the beat frequency $2\delta f=60\,$MHz used in the experiment. Strong minima appear when either of the optical frequencies is resonant with the atomic transition, since then the phase shift is close to zero and the loss is maximum. The optimum condition, with the optical frequencies symmetrically on either side of resonance, gives $\rm{FoM}\simeq 1/400$, indicating that $\sim400$ atoms in our $100\,\mu$m beam can be detected with a signal:noise ratio of 1.  The solid line shows FoM when we add a magnetic field of $0.65\,$mT, which is typical of the field experienced by atoms in our magnetic trap. This changes the frequency dependence of FoM quite significantly in the vicinity of the resonances. We see the two sharp minima associated with the dispersive character of $|\phi_{1}|$ move to higher frequency because they are due primarily to the $\sigma^+$ transition. In addition, we see two broader minima that are down-shifted. These are due to the $\sigma^-$ resonances that maximise the loss. This separation of the $\sigma^{+}$ and $\sigma^{-}$ transitions makes it possible to double the highest figure of merit. When detecting with parallel polarisation, the loss is more severe, as we have discussed in Sec.~\ref{sec:loss}. An addition, one cannot use the Zeeman shift to improve the figure of merit since the same transition produces both the phase shift and the loss.

The figure of merit is a convenient way to explore the application of the technique to other circumstances. For example in a Bose-Einstein condensate (BEC), any spontaneous emission will remove an atom from the condensate, so the $\sigma^+$ excitations all contribute to the loss. This typically increases the value of $q$ in Eq.~(\ref{eq:FoM}) by a factor of 16 because of the $15:1$ ratio of coupling strengths shown in Fig.~\ref{fig:transitions}. The condensate can be addressed on an atom chip by a beam of small waist $\sim2\,\rm{\mu m}$~\cite{kohnen11}, which enhances FoM by a factor of 50 in comparison with the $100\,\rm{\mu m}$ beam used here. The net effect is a FoM very similar to the dashed curve in Fig.~\ref{eq:FoM} but with a peak at the centre of $~0.03$. With a typical $^{87}$Rb condensate of $3\times 10^4$ atoms in prolate trap of trapping frequencies $20\,$Hz and $1\,$kHz, $10^3$ atoms are illuminated when the probe beam is centred on the cloud. Thus, the figure of merit shows that with our method, a 10\% measurement of the central density entails to loss of only 100 atoms from the condensate. The passage of a light pulse through a BEC normally imposes a local phase shift on the condensate equal to $\frac{N_{\gamma}}{N_{a}}\phi$. The gradient of this phase corresponds to a force that can excite phonons in the BEC. However, the two frequency components of the probe pulse used in this case induce opposite forces, which are equal in magnitude because they are symmetrically disposed on either side of resonance. As a result, the BEC will be undisturbed apart from the noise in the force due to photon statistics.

\section{Conclusion}

In conclusion, this paper presents a method of measuring the column density through a small part of an atom cloud    with minimum disturbance of the atoms. The method is to measure the phase shift between two synthesised frequency components of a laser beam, tuned on opposite sides of an atomic resonance. We have used this scheme to measure the column density of a magnetically trapped atom cloud and to monitor oscillations of the cloud in real time. Measurement sensitivity is principally limited by photon shot noise and excess noise due to the avalanche amplification in the photodiode. We have measured how many atoms are lost from the trap as an unwanted byproduct of the measurement. We have developed a figure of merit for this scheme, which quantifies the relationship between the sensitivity and the destructiveness of the measurements. Using this we have anticipated the performance of the technique when applied to Bose-Einstein condensates trapped on an atom chip.

We acknowledge the technical expertise of Valerijus Gerulis, who built the demodulation electronics and the support and advice of John Dyne and Stephen Maine in constructing the apparatus. We thank Tim Cable from Analog Modules for making sure we could use the last ever photo detector of the 712A-4 series. This research was supported by EPSRC (UK), the Royal Society (UK) and European projects AQUTE and HIP (EU).

\bibliographystyle{unsrt}
\bibliography{beat_note}

\end{document}